# Technical Note: Does the greater power of pencil beam scanning reduce the need for a proton gantry? A study of head-and-neck and brain tumors

Susu Yan[1,4], Nicolas Depauw[1], Judith Adams[1], Bram L. Gorissen[1], Helen A. Shih[2], Jay Flanz[1], Thomas Bortfeld[1], Hsiao-Ming Lu[3]

1. Division of Radiation Biophysics, Department of Radiation Oncology, Massachusetts General Hospital and Harvard Medical School
2. Department of Radiation Oncology, Massachusetts General Hospital and Harvard Medical School
3. Hefei Ion Medical Center and Ion Medical Research Institute, University of Science and Technology of China, Hefei, China
4. Corresponding author, syan5@mgh.harvard.edu.

## Abstract

**Purpose:** proton therapy systems without a gantry can be more compact and less expensive in terms of capital cost, and therefore more available to a larger patient population. Would the advances in pencil beam scanning and robotics make gantry-less treatment possible? In this study, we explore if high-quality treatment plans can be obtained without a gantry.

**Methods and Materials:** we recently showed that proton treatments with the patient in an upright position may be feasible with a new soft robotic immobilization device and imaging which enables multiple possible patient orientations during a treatment. In this study, we evaluate if this new treatment geometry could enable high quality treatment plans without a gantry. We created pencil beam scanning (PBS) treatment plans for seven patients with head-and-neck or brain tumors. Each patient was planned with two scenarios: one with a gantry with the patient in supine position and the other with a gantry-less fixed horizontal beam-line with the patient sitting upright. For the treatment plans, dose-volume-histograms (DVHs), target homogeneity index (HI), mean dose, $D_2$ and $D_{98}$ are reported. A robustness analysis of one plan was performed with $\pm$2.5 mm setup errors and $\pm$3.5% range uncertainties with nine scenarios.

**Results:** most of the PBS-gantry-less plans had similar target HI and OAR mean dose as compared to PBS-gantry plans, and similar robustness with respect to range uncertainties and setup errors.

**Conclusions:** pencil beam scanning provides sufficient power to deliver high quality treatment plans without requiring a gantry for head-and-neck or brain tumors. In combination with the development of the new positioning and immobilization methods required to support this treatment geometry, this work suggests the feasibility of further development of a compact proton therapy system with a fixed horizontal beam-line to treat patients in sitting and reclined positions.

## Introduction

Charged particle therapy is generally considered as the most precise form of radiation therapy. Proton treatment techniques have evolved from double scattering (DS) to pencil beam scanning (PBS), which is the standard delivery technique nowadays because of its capability to provide dose modulated proton therapy with superior dose distributions and requiring fewer fields than intensity modulated radiation therapy (IMRT). Studies have shown that PBS has lower dose to organs-at-risk (OARs) and similar or superior target coverage compared to IMRT for many treatment sites [1-4]. In the last years, PBS systems have further improved and now provide very precise delivery with a sharp penumbra.

Historically, patients were treated using DS, and the gantry played a critical role. For example, non-coplanar beams are generally necessary with DS to cover the target while avoiding OARs. With PBS, plans are less strongly dependent on beam angles because of the flexibility inherent to dose modulation in three dimensions. Studies have shown that PBS has the benefit of increased normal tissue sparing and better target conformity compared to DS treatment [4-9]. Several studies have investigated PBS and DS plan qualities for different treatment sites. PBS provides the opportunity to dispense with the patching technique of DS for the paraspinal and skull base sites [5]. One investigation assessed sixteen cases covering four tumor sites and suggested that PBS with manually optimized beam angles was effective for reducing the doses to some OARs as compared to DS plans [6]. A study which focused on non-small-cell lung cancer used two to four beams with PBS, which also reduced the dose to normal tissues relative to DS plans [4]. Single field uniform dose (SFUD) PBS for postmastectomy treatment has also shown dosimetric advantages to DS [8]. PBS has become the standard of practice now for newly constructed proton centers.

Is the dose modulation the only advantage for PBS? Could this also provide the advantage of removing the need for proton gantry and thereby reduce facility cost and increase the accessibility of proton therapy to patients?

Proton therapy started with fixed beam-lines several decades ago. These treatments were initially made possible with wood or styrofoam blocks and a wooden chair [9]. With the development of the gantry, more proton centers were developed and have been treating patients with the DS technique with multiple rooms. Fixed horizontal beam-lines since have been used to treat eye tumors almost exclusively for decades. A few centers have made efforts on treating other tumors without a gantry, such as brain tumor treatment using a chair [10] or articulating couch [11], with DS non-coplanar beams. Current proton therapy clinics are mostly equipped with a gantry or a half gantry with PBS. Even though there are 112 proton centers worldwide [12], proton therapy is still a limited resource compared to more than 8000 photon clinics in the world. The higher capital cost of proton therapy is one of the factors limiting its broader application [13, 14]. Currently, less than 1% of patients who can benefit from proton therapy receive it [15].

With modern PBS technology and the developments in robotics and imaging, we hypothesize that a gantry is no longer necessary for the proton therapy system. Without a gantry, the cost of proton centers would be substantially less which in turn could increase the availability of proton therapy. Our previous study showed that most patients could have been treated, geometrically, without a gantry based on 10 years proton patient treatment data [16]. In addition, we developed a first prototype of a robotic patient positioning chair with a soft robot immobilization device [17]. This system consists of a chair mounted on a robotic positioner, and a soft robotic immobilization device that conforms to the patient's body. The preliminary tests of the system showed reduced motion of healthy volunteers, especially the slouching movement in the sitting position, using the soft robot immobilization [17].

Figure 1 shows an example of the proposed gantry-less compact proton therapy system. The patient positioner and the imaging system are in the center of a synchrotron with a horizontal beam-line. The patient is sitting upright, such that coplanar beams can be delivered by rotating the chair.

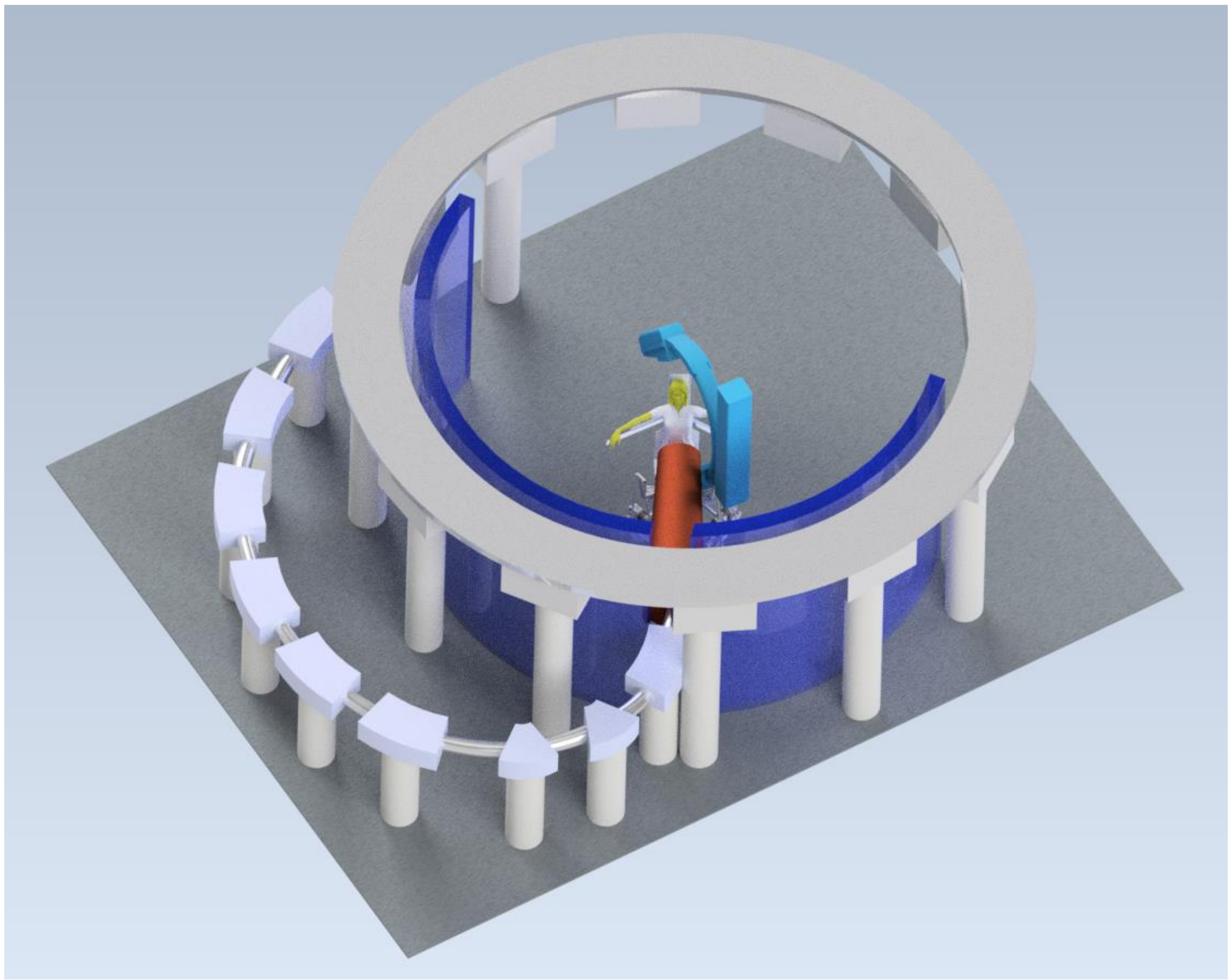

**Figure 1: 3-dimension model of a compact gantry-less single room proton system with a patient positioner and imaging system. The dimension of the simulated treatment room is 7x7 m$^2$. The patient is positioned on a reclinable chair. (3D model by Fernando Hueso-González).**

The main geometric difference between the gantry-less system and traditional gantry system is that gantries support beam delivery from both coplanar and non-coplanar directions. For DS delivery, non-coplanar angles were necessary for high quality treatment plans. They were used to avoid OARs close to the target and to make the air gap uniform between the aperture and the skull, thereby reducing heterogeneities along the beam path. However, with a modern PBS delivery system with a small spot size, non-coplanar beam directions may not be essential, because the dose can be modulated at each spot position.

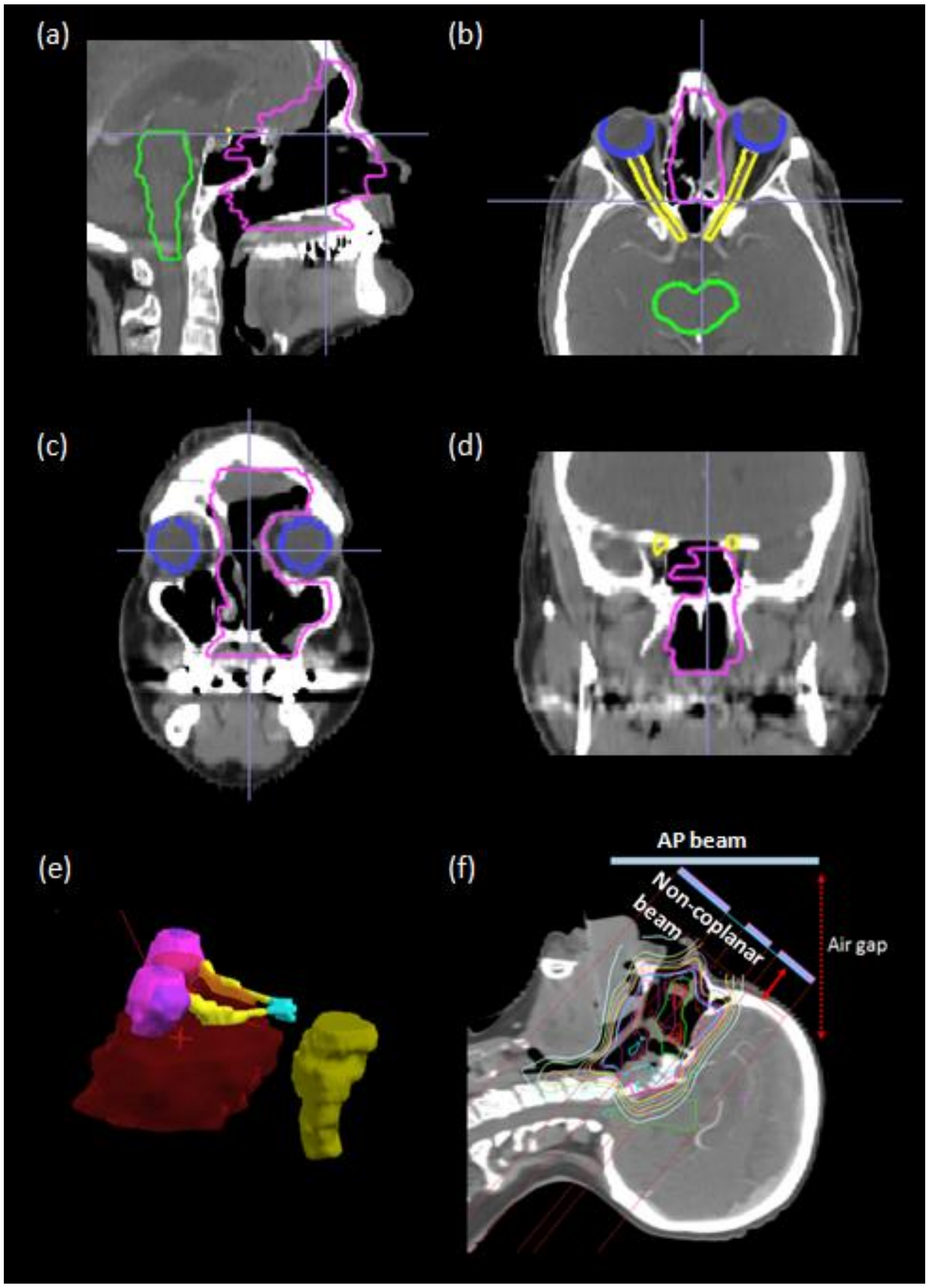

**Figure 2: Target and OAR for patient 2 with a complex paranasal tumor. (a) Sagittal view (b) axial view, (c) and (d) coronal view of the target (pink) which is close to the optic nerves (yellow), eyes (blue), and brain stem (green). (e) 3D view of the target (red) and OARs: eyes (purple), optic nerves (yellow), chiasm (cyan) and brain stem (dark yellow). (f) A non-coplanar beam is chosen instead of an AP beam to minimize the air gap between the aperture and the target for uniform dose.**

For one paranasal tumor case, shown in Figure 2, the target is close to the chiasm, brainstem, optic nerves, eyes and left retina. The DS technique uses the spread-out Bragg peak (SOBP) where the modulation width is equal to the maximum length of the target dimension along the beam path. In contrast, the PBS technique provides a three-dimensional (3D) dose modulation at each beam spot position in the target. The differences between DS delivery and 3D modulation play an important role in beam angle selection for treatment. The lack of control over the shape of the depth-dose curve makes DS more dependent on the beam angle to minimize dose to OARs. Additionally, Figure 2 (f) shows that an anterior-posterior (AP) beam would have a non-uniform air gap between the beam and the curved skull. This could cause hot or cold spots in the target region with DS delivery. Therefore, a non-coplanar DS beam (anterior oblique) is needed which is enface with a more uniform air gap. In contrast, this non-uniform air gap does not affect PBS because the range of the beam spot at each position in the target is modulated so that the entire plan delivers a uniform dose.

In this proof-of-concept study, we investigate if gantry-less treatment plans using modern PBS and only coplanar beams, and with appropriate immobilization and imaging, could achieve the same dosimetric goals as plans generated with the full flexibility of a gantry.

## Materials and Methods

### 1. Patient selection for treatment planning

Seven patients who were treated in the supine position with the gantry at our proton center were selected. Table 1 summarizes the clinical information for these patients, including disease site, prescription, constraints to OARs, beam angles and patient positions. These patients have tumors in the head-and-neck or brain region. Three of the seven patients had paranasal tumors, which were considered among the most challenging cases to plan without a gantry, as discussed in the previous session. The fourth patient's treatment included nasal cavity and neck nodes, which was also deemed to be challenging to plan without utilization of non-coplanar beam options. The fifth patient had a suprasellar tumor, for which multiple non-coplanar beams are generally used to mitigate potential heterogeneous end range radiobiological effectiveness increased with fewer beams. The last two of the seven patients had a frontal lobe astrocytoma and a clival chordoma. The OAR constraints were prescribed by the physician for each individual case.

### 2. Treatment planning

Two PBS plans were generated for each patient, shown in the third column of Table 1:
1) A PBS-gantry-less plan that can be delivered without the gantry with the patient in a sitting position, which means that only coplanar beams are used. The coplanar beams were mostly chosen as a subset of the beam angles used in the PBS-gantry plans for simplicity of the comparison.
2) A traditional PBS-gantry plan with the patient in supine positions, which contains non-coplanar beams. The PBS plans were generated using the in-house developed ASTROID software [20]. The PBS plans were optimized for multiple fields (MFO) or single field optimization (SFO) with multi-criteria optimization (MCO) and Pareto-surface navigation [21]. These plans were optimized to deliver a uniform dose based on the full prescription as a composite of the beams. Proton plans typically do not have a planning target volume (PTV), because the dose distribution is highly sensitive to variations of anatomies in the beam path. Prescriptions are made to clinical target volume (CTV) or gross tumor volume (GTV). SFO plans were optimized such that each beam was constrained by the prescription divided by the number of beams. When the clinical gantry plan was optimized with SFO, the gantry-less plan was also generated using SFO. Otherwise, the plan was optimized using MFO. After the optimization, MCO plans were then navigated to the most desirable plan. The spot size in air ranges from 2.5 mm to 4.5 mm (σ) at isocenter with energies from 240 MeV down to 70 MeV. Modern proton centers produce beams with this spot size, e.g. Paul Scherrer Institute (PSI)'s PBS system [22] and the Gordon Browne proton center at Massachusetts General Hospital.

**Table 1. The indication, prescription and constraints for seven patients treated with DS and PBS techniques, together with the table and gantry angles used in the PBS-gantry-less plan and PBS gantry plan. Planning complexity was determined by a proton treatment planning expert.**

| **Patient (Planning complexity)** | **Prescription dose, requested coverage, fractions and constraints** | **Number of beams in PBS-gantry-less** (equivalent couch angle, equivalent gantry angle), treatment position<br>**Number of beams in PBS-gantry** (couch angle, gantry angle), treatment position |
|---|---|---|
| **1 Paranasal (High)**<br>(Recurrent low grade chondrosarcoma) | **72 Gy(RBE), 100% to GTV, 40 fx**<br>Brainstem < 55 Gy(RBE)<br>Chiasm < 62 Gy(RBE)<br>Optic nerve <62 Gy(RBE)<br>Retina <62 Gy(RBE) | PBS-gantry-less MFO: 3 beams<br>(0,0) (0,90) (0,270), **Upright sitting in a chair**<br><br>PBS-gantry MFO: 5 beams (2 beams need a gantry)<br>(0,0) (0,90) (0,270) (270,10) (80,300), **Supine** |
| **2 Paranasal (High)**<br>(Complex CTV involving the nasal cavity, left maxillary, ethmoid and frontal sinuses, very close to the left optic nerve and medial eye) | **60 Gy(RBE), 95% to CTV, 30fx**<br>Brainstem < 54 Gy(RBE)<br>Left retina < 50 Gy(RBE)<br>Left optic nerve < 60 Gy(RBE)<br>Right optic nerve < 54 Gy(RBE) | PBS-gantry-less MFO: 3 beams<br>(0,0) (0,90) (0,270), **Upright sitting in a chair**<br><br>PBS-gantry MFO: 5 beams (2 beams need a gantry)<br>(0,0) (0,90) (0,270) (270,15) (80,300), **Supine** |
| **3 Paranasal Sinus retreatment (High)** | **59.4 Gy(RBE), 100% to CTV, 33 fx**<br>Brainstem < 20 Gy(RBE)<br>Chiasm < 20 Gy(RBE)<br>Optic nerves < 24 Gy(RBE) | PBS-gantry-less MFO: 2 beams<br>(0,90) (0,270), **Upright sitting in a chair**<br><br>PBS-gantry MFO: 5 beams (1 beam needs a gantry)<br>(0,100) (0,260) (0,90) (0,270) (90,310), **Supine** |
| **4 Nasal cavity with neck nodes (High)** | **CTV combine 60 Gy(RBE), 100% to CTV, 30 fx**<br>**GTV boost, 66 Gy(RBE), 100% to GTV, 30 fx**<br>Optic nerves <54-56 Gy(RBE)<br>Chiasm < 54 Gy(RBE)<br>Brainstem < 54 Gy(RBE)<br>Spinal cord < 45 Gy(RBE) | PBS-gantry-less MFO: 3 beams<br>(0,0) (0,70) (0,290), **Upright sitting in a chair**<br><br>PBS-gantry MFO: 4 beams (1 beam needs a gantry)<br>(0,0) (0,70) (0,290), (90,315), **Supine** |
| **5 Brain (Low)** (Suprasellar) | **GTV 52.2 Gy(RBE), 100% to GTV, 29 fx**<br>Brainstem < 54 Gy(RBE)<br>Optic nerves <54 Gy(RBE)<br>Chiasm < 54 Gy(RBE)<br>Hippocampus < 45 Gy(RBE) | PBS-gantry-less MFO: 2 beams<br>(0,90) (0,270), **Upright sitting in a chair**<br><br>PBS-gantry MFO: 3 beams (1 beam needs a gantry)<br>(0,90) (0,270), (90,266), **Supine** |
| **6 CNS (Medium)**<br>(Astrocytoma of frontal lobe) | **CTV= 59.4 Gy(RBE), 100% to CTV, 33 fx**<br>Optic nerves < 54 Gy(RBE)<br>Chiasm < 54 Gy(RBE)<br>Eyes < 45 Gy(RBE)<br>Lacrimals < mean 26 Gy(RBE)<br>R hippocampus: ALARA<br>Pituitary: ALARA | PBS-gantry-less SFO: 2 beams<br>(0,245) (0,275), **Upright sitting in a chair**<br><br>PBS-gantry SFO: 4 beams (2 beams need a gantry)<br>(0,275) (0,340) (5,240) (90,305), **Supine** |
| **7 Chordoma (High)**<br>(Chordoma of clivus) | **GTV 72.2 Gy(RBE), 100% to GTV**<br>**CTV 68.4 Gy(RBE), 100% to CTV, 38 fx**<br>Brainstem < 55 Gy(RBE)<br>Optic nerves < 62 Gy(RBE)<br>Cochlea: - ALARA | PBS-gantry-less SFO: 2 beams<br>(0,300) (0,60), **Upright sitting in a chair**<br><br>PBS-gantry SFO: 6 beams (1 beam need a gantry)<br>(0,180) (0,300) (0,60) (90,60) (0,225) (0,140), **Supine** |

Dose-volume-histograms (DVHs) were calculated for all the plans for each patient. For the target region, a homogeneity index (HI) was calculated based on equation 1,

$$HI = \frac{D_2 - D_{98}}{D_P} \times 100\% \tag{1}$$

where $D_2$ is the near-maximum dose (only 2% of the voxels receive higher dose), $D_{98}$ is the near-minimum dose (98% of voxels receive higher dose), and $D_P$ is the prescription dose. HI equals zero for a perfectly homogeneous dose, and is positive for inhomogeneous dose. The relative volume covered by the 95% isodose ($V_{95}$) is also reported for the target. In addition, the mean dose ($D_{\mathrm{mean}}$), $D_2$ and $D_{98}$ of both the target and OARs are reported. A robustness analysis with $\pm$2.5 mm setup errors and $\pm$3.5% range uncertainties was performed for the PBS-gantry-less and PBS-gantry plans for paranasal patient 2 [31]. The analysis for setup errors was performed by moving the patient in the planning system in lateral, anterior-posterior and superior-inferior directions by $\pm$2.5mm. Range uncertainties were simulated by rescaling the proton relative stopping power ratio map by $\pm$3.5% to change the proton range.

**Results**

For all seven patients, PBS-gantry plans and PBS-gantry-less plans were generated. Figure 3 shows the dose distribution of these two plans for patients 1 and 2. The DVHs of these plans are shown in Figure 4 for comparison. In Figure 4(a)-(b), PBS-gantry-less and gantry plan have almost the same DVHs, except for the slightly reduced dose to the optic nerves for paranasal patient 2 with the PBS-gantry plan. For paranasal patient 3 in Figure 4(c), the PBS-gantry plan compared to the PBS-gantry-less plan has improved target coverage and OARs sparing, although the dose to OARs is relatively low already. For the nasal cavity with neck nodes patient in Figure 4(d), there is a minimal difference between the two PBS plans with and without a gantry. For the patient with suprasellar tumor in Figure 4(e), the target coverage is similar, while the PBS-gantry plan reduces the dose to right and left hippocampi by 31% and 16%, respectively. All other OARs have similar dose. For the CNS patient in Figure 4(f), the PBS-gantry plan with 4 beams slightly improves target coverage. The PBS-gantry-less plan reduces the dose to right hippocampus, brainstem and right lacrimal compared to the PBS-gantry plan. For the chordoma patient in Figure 4(g), the PBS-gantry-less plan reduces dose to brainstem and chiasm compared to the PBS-gantry plan, while the dose to other OARs and target coverage are similar. Table 2 shows the quantitative results of the DVHs for the PBS-gantry and PBS-gantry-less plans.

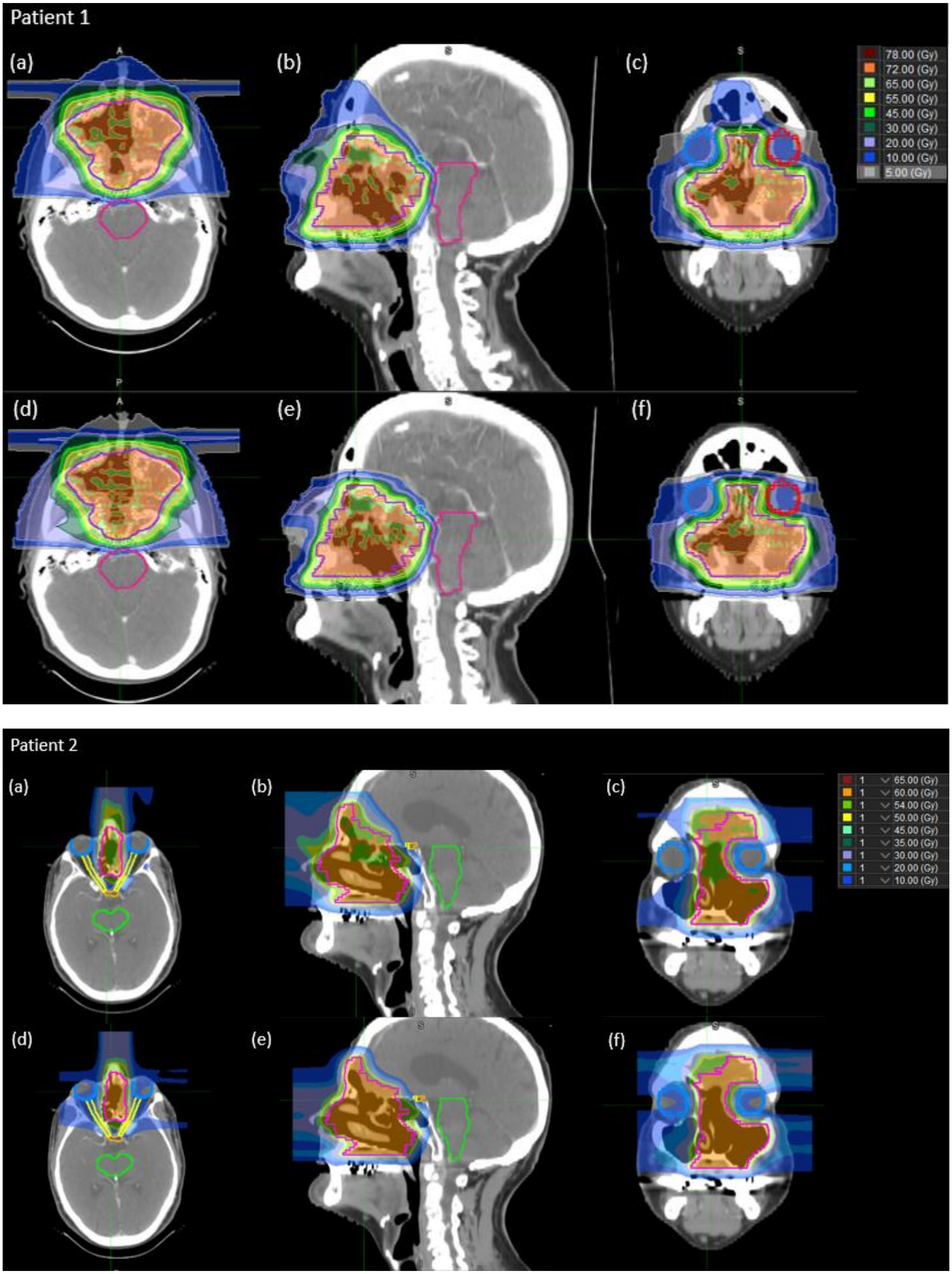

**Figure 3: Dose distribution of PBS-gantry plan (a)-(c) and PBS-gantry-less plan (d)-(f) for patient 1 with paranasal tumor, and patient 2 with complex paranasal tumor.**

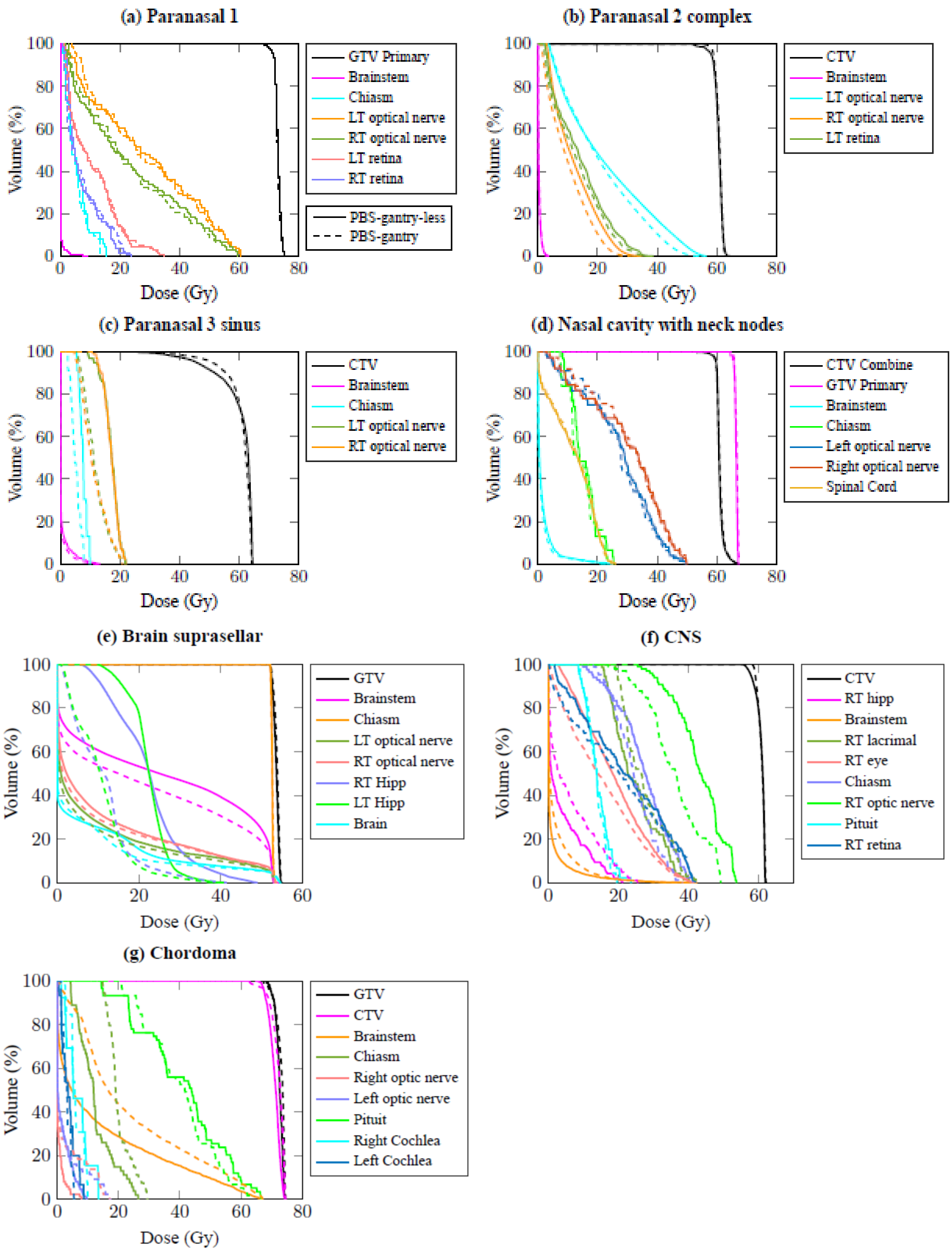

**Figure 4: (a-g) Dose-volume-histograms (DVHs) of patients 1-7 for pencil beam scanning (PBS)-gantry-less plans (solid line) and PBS-gantry plans (dashed line).**

**Table 2: HI, $V_{95}$, $D_{mean}$, $D_2$ and $D_{98}$ of target, and $D_{mean}$, $D_2$, $D_{98}$ of OARs for the PBS-gantry-less and PBS-gantry plans.**

| Case number and target site | | PBS-gantry-less | | | | PBS-gantry | | | |
|---|---|---|---|---|---|---|---|---|---|
| **1 Paranasal** | **Target** | **HI (%)** | **$V_{95}$ (%)** | **$D_2$ (Gy)** | **$D_{98}$ (Gy)** | **HI (%)** | **$V_{95}$ (%)** | **$D_2$ (Gy)** | **$D_{98}$ (Gy)** |
| | | 6.4 | 100 | 74.9 | 70.4 | 6.4 | 100 | 74.9 | 70.2 |
| | **OAR** | **Dmean (Gy)** | | **$D_2$ (Gy)** | **$D_{98}$ (Gy)** | **Dmean (Gy)** | | **$D_2$ (Gy)** | **$D_{98}$ (Gy)** |
| | **LT optical nerve** | 25.4 | | 59.5 | 2.8 | 24.9 | | 60.2 | 2.6 |
| | **RT optical nerve** | 21.5 | | 59.5 | 1.9 | 21.5 | | 59.5 | 2.6 |
| **2 Paranasal Complex** | **Target** | **HI (%)** | **$V_{95}$ (%)** | **$D_2$ (Gy)** | **$D_{98}$ (Gy)** | **HI (%)** | **$V_{95}$ (%)** | **$D_2$ (Gy)** | **$D_{98}$ (Gy)** |
| | | 7.9 | 98.6 | 62.2 | 57.5 | 7.6 | 99.3 | 62.3 | 57.7 |
| | **OAR** | **Dmean (Gy)** | | **$D_2$ (Gy)** | **$D_{98}$ (Gy)** | **Dmean (Gy)** | | **$D_2$ (Gy)** | **$D_{98}$ (Gy)** |
| | **LT optical nerve** | 24.0 | | 50.4 | 4.4 | 20.9 | | 45.8 | 4.1 |
| | **RT optical nerve** | 11.4 | | 27.9 | 1.4 | 10.0 | | 21.6 | 1.3 |
| **3 Paranasal Sinus** | **Target** | **HI (%)** | **$V_{95}$ (%)** | **$D_2$ (Gy)** | **$D_{98}$ (Gy)** | **HI (%)** | **$V_{95}$ (%)** | **$D_2$ (Gy)** | **$D_{98}$ (Gy)** |
| | | 44.3 | 84.1 | 64.4 | 38.1 | 34.6 | 88.9 | 64.2 | 43.6 |
| | **OAR** | **Dmean (Gy)** | | **$D_2$ (Gy)** | **$D_{98}$ (Gy)** | **Dmean (Gy)** | | **$D_2$ (Gy)** | **$D_{98}$ (Gy)** |
| | **Brainstem** | 0.8 | | 12.4 | 0.0 | 0.5 | | 12.4 | 0.0 |
| | **Chiasm** | 8.9 | | 11.4 | 5.5 | 5.2 | | 8 | 2.5 |
| | **LT optical nerve** | 17.2 | | 21.9 | 12 | 11.3 | | 20.9 | 4.8 |
| | **RT optical nerve** | 16.9 | | 22 | 9.8 | 11.8 | | 21.8 | 5.9 |
| **4 Nasal cavity with neck nodes** | **Target** | **HI (%)** | **$V_{95}$ (%)** | **$D_2$ (Gy)** | **$D_{98}$ (Gy)** | **HI (%)** | **$V_{95}$ (%)** | **$D_2$ (Gy)** | **$D_{98}$ (Gy)** |
| | | 2.7 | 99.4 | 66.9 | 65.4 | 1.4 | 99.5 | 67.3 | 66.4 |
| | **OAR** | **Dmean (Gy)** | | **$D_2$ (Gy)** | **$D_{98}$ (Gy)** | **Dmean (Gy)** | | **$D_2$ (Gy)** | **$D_{98}$ (Gy)** |
| | **Chiasm** | 14.5 | | 25.1 | 8.2 | 14.2 | | 23.3 | 7.4 |
| | **LT optical nerve** | 27.7 | | 46.5 | 3.3 | 27.4 | | 48.4 | 2.9 |
| | **RT optical nerve** | 27.5 | | 49.6 | 3.2 | 30.7 | | 49.5 | 4.0 |
| **5 Brain** (Suprasellar) | **Target** | **HI (%)** | **$V_{95}$ (%)** | **$D_2$ (Gy)** | **$D_{98}$ (Gy)** | **HI (%)** | **$V_{95}$ (%)** | **$D_2$ (Gy)** | **$D_{98}$ (Gy)** |
| | | 4.7 | 100 | 53 | 50.5 | 4.8 | 100 | 54.9 | 52.4 |
| | **OAR** | **Dmean (Gy)** | | **$D_2$ (Gy)** | **$D_{98}$ (Gy)** | **Dmean (Gy)** | | **$D_2$ (Gy)** | **$D_{98}$ (Gy)** |
| | **Brainstem** | 26.1 | | 51.7 | 0.0 | 22.3 | | 52.9 | 0.0 |
| | **Chiasm** | 50.5 | | 51.3 | 52.1 | 52.6 | | 53.0 | 52.1 |
| | **RT Hippocampus** | 26.1 | | 44.6 | 7.6 | 10.7 | | 30.6 | 1.7 |
| | **LT Hippocampus** | 26.3 | | 39.9 | 11.9 | 11.2 | | 33.4 | 1.7 |
| | **Brain** | 7.9 | | 53.7 | 0.0 | 7.2 | | 53.9 | 0.0 |
| **6 CNS** (Astrocytoma) | **Target** | **HI (%)** | **$V_{95}$ (%)** | **$D_2$ (Gy)** | **$D_{98}$ (Gy)** | **HI (%)** | **$V_{95}$ (%)** | **$D_2$ (Gy)** | **$D_{98}$ (Gy)** |
| | | 8.4 | 98.8 | 61.9 | 56.8 | 5.5 | 99.3 | 62.0 | 58.7 |
| | **OAR** | **Dmean (Gy)** | | **$D_2$ (Gy)** | **$D_{98}$ (Gy)** | **Dmean (Gy)** | | **$D_2$ (Gy)** | **$D_{98}$ (Gy)** |
| | **RT Hippocampus** | 4.1 | | 20.4 | 0.0 | 6.5 | | 23.8 | 0.1 |
| | **Brainstem** | 1.4 | | 16.5 | 0.0 | 2.3 | | 18.2 | 0.0 |
| | **RT lachrymal** | 24.9 | | 38.6 | 15.7 | 27.4 | | 41.6 | 19.2 |
| | **RT eye** | 18.8 | | 38.5 | 3.4 | 15.5 | | 38.0 | 0.7 |
| | **Chiasm** | 27.2 | | 41.1 | 11.0 | 24.7 | | 36.2 | 13.7 |
| | **RT optic nerve** | 43.1 | | 53.4 | 26.6 | 36 | | 49.1 | 18.9 |
| | **Pituitary** | 14.0 | | 23.5 | 8.6 | 13.6 | | 19.5 | 8.0 |
| | **RT retina** | 22.0 | | 41.1 | 1.9 | 20.1 | | 41.1 | 0.3 |
| **7 Chordoma** (Chordoma of clivus) | **Target** | **HI (%)** | **$V_{95}$ (%)** | **$D_2$ (Gy)** | **$D_{98}$ (Gy)** | **HI (%)** | **$V_{95}$ (%)** | **$D_2$ (Gy)** | **$D_{98}$ (Gy)** |
| | | 6.4 | 95.4 | 74.1 | 69.3 | 7.6 | 99.1 | 74.4 | 68.7 |
| | **OAR** | **Dmean (Gy)** | | **$D_2$ (Gy)** | **$D_{98}$ (Gy)** | **Dmean (Gy)** | | **$D_2$ (Gy)** | **$D_{98}$ (Gy)** |
| | **Brainstem** | 14.9 | | 63.1 | 0.0 | 24.2 | | 65.4 | 1.1 |
| | **Chiasm** | 12.5 | | 25.6 | 4.4 | 20.2 | | 29.3 | 14.9 |
| | **RT optic nerve** | 0.7 | | 7.3 | 0.0 | 2.9 | | 16.8 | 0.0 |
| | **LT optic nerve** | 1.7 | | 8.8 | 0.0 | 2.7 | | 16.9 | 0.0 |
| | **Pituit** | 41 | | 66.3 | 14.5 | 40.4 | | 62.0 | 21.0 |
| | **RT cochlea** | 6.7 | | 13.5 | 1.3 | 6.2 | | 10.0 | 2.6 |
| | **LT cochlea** | 3.8 | | 8.7 | 0.5 | 3.1 | | 5.5 | 1.2 |

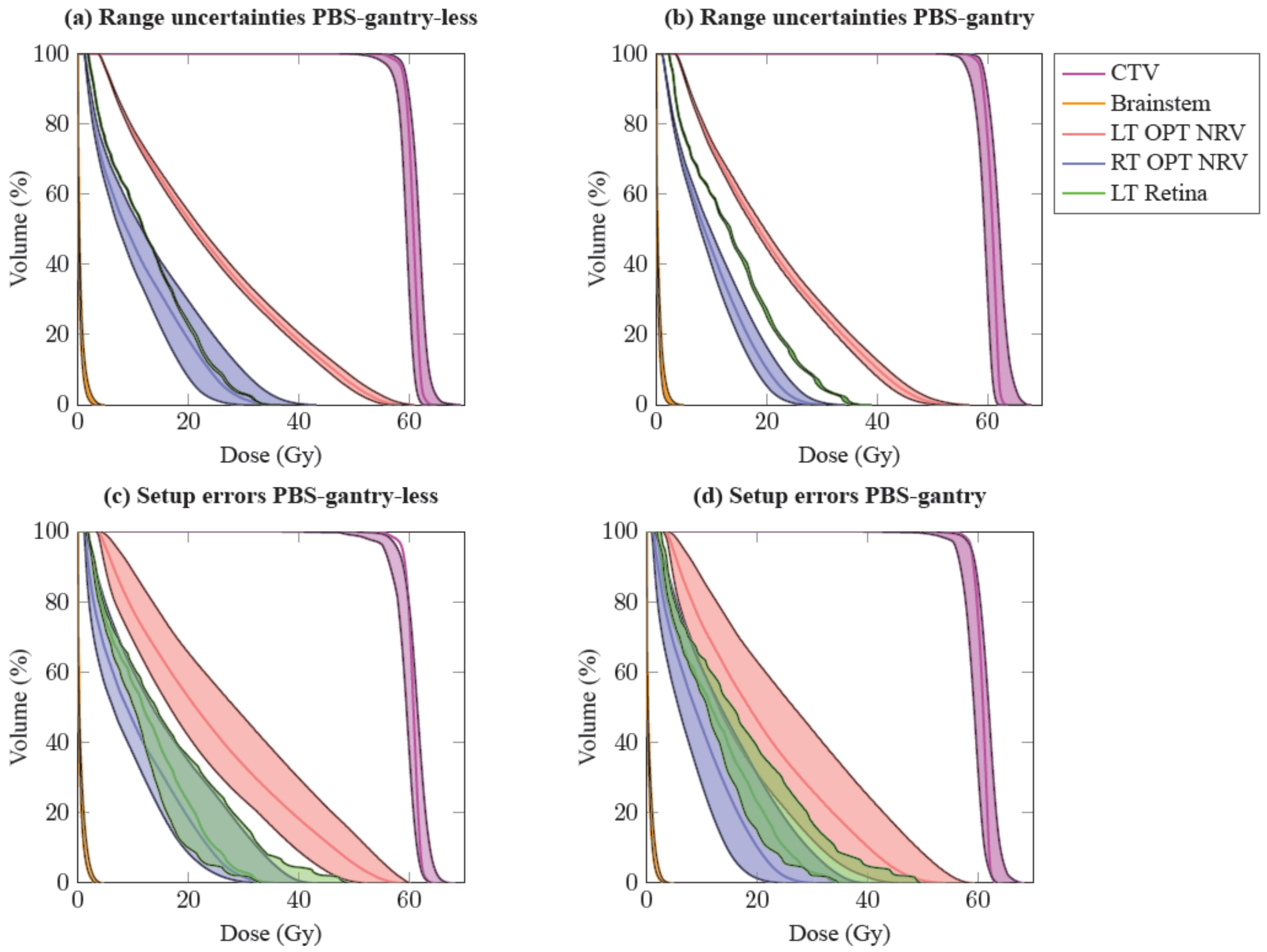

Figure 5: Robustness DVH envelopes for patient 2 with a paranasal tumor of the PBS-gantry-less and PBS-gantry plans. (a–b) DVH envelopes based on range uncertainties of 3.5%. (c–d) DVH envelopes based on setup errors of 2.5 mm. The nominal DVH is shown with solid colored curves.

The DVH envelopes with 3.5% range uncertainties and 2.5 mm setup errors in x-, y- and z-directions from the plan robustness evaluation for paranasal patient 2 are shown in Figure 5. The DVH envelopes are the maximum amplitude of the plans with different uncertainties and setup errors. The solid colored line in each DVH band is from the nominal plan. PBS-gantry-less and PBS-gantry have similar robustness. Both plans are more robust relative to the range uncertainties than setup errors. The constraints on the OARs are still achieved under range uncertainties and setup errors for both plans.

## Discussion and Conclusions

Our results show that when a modern PBS system is used, high-quality proton treatment plans can be obtained for brain and head-and-neck patients even when all non-coplanar beams are removed. This suggests that if appropriate positioning and immobilization is available, many proton treatments could be delivered using a gantry-less system that does not support non-coplanar fields.

Additionally, PBS-gantry-less and PBS-gantry plans were equally robust relative to range uncertainties and setup errors for one tested patient (Patient 2). Note here, these plans were generated with multi-criteria optimization but without robust optimization, which could diminish effects of uncertainties [24-27]. Robustness analysis is performed for each unique beam geometry for each specific treatment site in our clinic, but not for every patient. In some of the cases, such as patient 7, the gantry-less plan may have less robustness compared to the gantry plan because fewer beams used in the gantry-less plan. The two lateral anterior beams in the gantry-less plan range out to the brainstem. Compared to the six beams in the gantry plan, the mean dose to brainstem may increase if uncertainties are added. However, more coplanar beams could also be used to eliminate the uncertainties in the gantry-less geometry. Future studies using robust optimization are warranted. Robustness was evaluated against ±3.5% range errors. Range verification techniques [28-30] can reduce the magnitude of these errors, and further reduce their impact on plan quality.

A non-coplanar beam could have a small benefit for specific geometries. For example, for patients 3 and 4 (paranasal, nasal cavity and neck nodes), PBS-gantry plans had respectively 21.9% and 48.1% less target HI with non-coplanar beam angles of 50° and 45° compared to PBS-gantry-less plan. Another example, as shown in Figure 4(f) and Table 2, a right frontal lobe patient, the PBS gantry plan with four beams (two non-coplanar beams) reduces the mean dose to the right optic nerve, chiasm, right eye and right retina by 19.7%, 10%, 21.3%, 9.5% respectively compared to the PBS-gantry-less plan with two co-planar beams. The clinical significance of these slightly better PBS-gantry plans depends on clinical interpretation. Thus, the door is open to the discussion of the cost-benefit comparison for gantry-less systems. Although it is desirable to reduce the dose to OARs as much as reasonably achievable, an analysis of including more beams, which extend both the irradiation time and the set-up time, could be useful. Our gantry-less plans used on average only 45% of the number of beams of the gantry plans. The reduced irradiation and set-up time from gantry-less system would be expected to increase treatment robustness and shorten overall treatment time.

The beam angles used in this study for the PBS-gantry-less plan were a subset of the beam angles used in the PBS-gantry plans for simplicity of the comparison. Different or more co-planar beam angles could likely improve the OAR doses. For example, additional co-planar beams could be used for complex cases with targets near multiple OARs. Beam angle optimization has been widely studied in photon therapy and adapted to PBS [23]. Possible future studies could adapt beam angle optimization to planning for fixed beam-line planning.

A limitation of this study is the limited number of patients and disease sites that were explored. Statistical analysis on more patients in a future study is warranted. The results encourage us to further investigate more patients with generalized disease sites to create disease site specific guidelines on beam angles for fixed beam-line treatment, as well as further development of gantry-less proton therapy solutions. Many head-and-neck patients could be treated in an upright position with only co-planar beams. Certain non-coplanar beams could also be achieved by tilting and rotating the patient on a chair or bending the beam by a small angle. While a gantry-less system could sometimes require changing the treatment position from the supine orientation that is the current standard in radiotherapy to other patient orientations, new immobilization and imaging systems can be used to meet these challenges. Recent developments in robotics and imaging may enable treatments in orientations that were previously considered infeasible or too difficult. We are developing a new device that combines patient immobilization based on soft robotics with a 6-degree of freedom positioner [17]. Despite the positioning challenges, a gantry-less system has a much simpler beam transportation which could provide smaller a spot size compared to a gantry system. This advance could improve the treatment plan quality with less beams.

In X-ray therapy, several new treatment devices have been introduced in the last years that changed from the traditional C-arm LINAC geometry to an enclosed bore geometry with only co-planar beams, largely driven by developments in the beam delivery technique. These devices offer advances such as faster treatments, additional imaging modalities, and reduced room size. Similarly, modern proton PBS delivery and recent technology developments could facilitate rethinking of the traditional gantry-based proton delivery geometry. The potential reduction in cost and size of charged particle therapy systems may be great and the accessibility gains for patients who can benefit from the improved precision of proton therapy may be even greater.